\documentclass[prl,amsmath,amssymb,superscriptaddress,twocolumn,longbibliography]{revtex4}
\usepackage{dcolumn}
\usepackage{bm}
\usepackage{bbm}
\usepackage{amssymb, amsbsy, amsmath, latexsym, dsfont, array, layout, graphicx,mathrsfs,color}
\usepackage{makecell,multirow}
\usepackage[colorlinks,linkcolor=blue,anchorcolor=red,citecolor=red]{hyperref}

\usepackage{array}
\newcommand{\PreserveBackslash}[1]{\let\temp=\\#1\let\\=\temp}
\newcolumntype{C}[1]{>{\PreserveBackslash\centering}p{#1}}
\newcolumntype{R}[1]{>{\PreserveBackslash\raggedleft}p{#1}}
\newcolumntype{L}[1]{>{\PreserveBackslash\raggedright}p{#1}}

\begin{document}

\title{Nonclassical oscillations in pre- and post-selected quantum walks}

\author{Xiao-Xiao Chen}
\affiliation{Center for Quantum Technology Research and Key Laboratory of Advanced Optoelectronic Quantum Architecture
and Measurements (MOE), School of physics, Beijing Institute of Technology, Haidian District, Beijing 100081, People's Republic of China}
\author{Zhe Meng}
\affiliation{Center for Quantum Technology Research and Key Laboratory of Advanced Optoelectronic Quantum Architecture
and Measurements (MOE), School of physics, Beijing Institute of Technology, Haidian District, Beijing 100081, People's Republic of China}
\author{Jian Li}
\affiliation{Center for Quantum Technology Research and Key Laboratory of Advanced Optoelectronic Quantum Architecture
and Measurements (MOE), School of physics, Beijing Institute of Technology, Haidian District, Beijing 100081, People's Republic of China}
\author{Jia-Zhi Yang}
\affiliation{Center for Quantum Technology Research and Key Laboratory of Advanced Optoelectronic Quantum Architecture
and Measurements (MOE), School of physics, Beijing Institute of Technology, Haidian District, Beijing 100081, People's Republic of China}
\author{An-Ning Zhang}\email{anningzhang@bit.edu.cn}
\affiliation{Center for Quantum Technology Research and Key Laboratory of Advanced Optoelectronic Quantum Architecture
and Measurements (MOE), School of physics, Beijing Institute of Technology, Haidian District, Beijing 100081, People's Republic of China}
\author{Tomasz Kopyciuk}
\affiliation{Faculty of Physics, Adam Mickiewicz University, Uniwersytetu Pozna{\'n}skiego 2, 61-614 Pozna\'{n}, Poland}
\author{Pawe{\l} Kurzy\'{n}ski}\email{pawel.kurzynski@amu.edu.pl}
\affiliation{Faculty of Physics, Adam Mickiewicz University, Uniwersytetu Pozna{\'n}skiego 2, 61-614 Pozna\'{n}, Poland}

\date{\today}

\begin{abstract}
Quantum walks are counterparts of classical random walks. They spread faster, which can be exploited in information processing tasks, and constitute a versatile simulation platform for many quantum systems. Yet, some of their properties can be emulated with classical light. This rises a question: which aspects of the model are truly nonclassical? We address it by carrying out a photonic experiment based on a pre- and post-selection paradox. The paradox implies that if somebody could choose to ask, either if the particle is at position $x=0$ at even time steps, or at position $x=d$ ($d>1$) at odd time steps, the answer would be positive, no matter the question asked. Therefore, the particle seems to undergo long distance oscillations despite the fact that the model allows to jump one position at a time. We translate this paradox into a Bell-like inequality and experimentally confirm its violation up to eight standard deviations.
\end{abstract}

\maketitle


\section{Introduction}

We study non-classicality of a discrete-time quantum walk (DTQW) \cite{Aharonov,Meyer}, a quantum counterpart of the quintessential random walk, in which the walker and the driving coin are quantum systems capable of becoming superposed and interfering. The interference is responsible for a behavior that strikingly differs from the classical random diffusion. DTQWs spread ballistically faster and their spatial probability distribution is far from Gaussian.

Despite the fact that DTQWs can simulate various quantum systems and their ballistic spreading properties are used in a number of information processing algorithms \cite{Review1,Review2,Review3,Review4}, it is possible to emulate their behavior with classical light \cite{CQW1,CQW2,CQW3,CQW4}. More precisely, if instead of a particle, such as a photon, one used in a DTQW experiment a classical coherent light beam, the beam's  amplitude would mimic the walker's probability amplitude. Does it mean that there is nothing nonclassical about DTQWs? The only difference between the two scenarios seems to lie in the measurement. In case of the classical light one performs a single experiment that yields intensity distribution at all positions. In the quantum case one performs many experiments, each resulting in a single click at a random position, and only later one evaluates a probability distribution that matches the classical intensity pattern. Therefore, in order to expose any nonclassicality in DTQWs, one should focus on measurements and examine if some of their properties lack a classical description.

Lack of classical description means that measurements on a system cannot be described by a particular type of hidden variables (HVs). HVs provide a classical probabilistic description of measurements \cite{Fine} under some physically motivated assumptions. The most common physically motivated assumptions are locality \cite{Bell} (a measurements in one location does not affect a measurement in some other, spatially separated location), non-contextuality \cite{KS} (an outcome of one measurement does not depend on which other compatible measurement is performed together with it) and so called macro-realism \cite{LG} (a measurement at time $t_0$ does not influence the outcome of a measurement at some later time $t_1$).

In \cite{QWLG} it was experimentally confirmed that DTQWs do not meet the macro-realism assumption. Here, we experimentally confirm that DTQWs do not meet the non-contextuality assumption. We do this by designing a Bell-like inequality \cite{Bell} and showing that experimentally obtained measurement data violates it. The inequality is based on a recent logical pre- and post-selection (LPPS) paradox designed by some of the authors \cite{UsNJP}. Notice that LPPS paradoxes were shown to be proofs of contextuality \cite{PPSc1,PPSc2}, hence if some system admits such a paradox, it is nonclassical in a sense that it is contextual \cite{KS}.


\section{Results}


\subsection{Model}

We consider a DTQW on a 1D lattice. The system's state is given by $|x\rangle \otimes |c\rangle$, where $x\in {\mathbb{Z}}$ is the position of the walker and $c=\pm$ represents two states of the coin. A single step is described by an operator $U=S(\openone\otimes C)$, where $S$ is the conditional translation
\begin{equation}
S|x\rangle\otimes|\pm\rangle = |x\pm 1\rangle\otimes|\pm\rangle
\end{equation}
and $C$ is the coin-toss operation that we choose to be $C=NOT$, i.e., $C|\pm\rangle = |\mp\rangle$.

The above evolution is periodic and the period is just two steps. A state $|x\rangle\otimes|\pm\rangle$ after the first step becomes $|x \mp 1\rangle\otimes|\mp\rangle$, but after the second step it returns to $|x\rangle\otimes|\pm\rangle$. Due to reasons that will be explained in a moment, we are interested in an initial state that is a particular superposition of position and coin states. We are going to initialize the system (pre-select it) in the state
\begin{equation}\label{pre0}
|pre(0)\rangle = \frac{1}{\sqrt{5}}\left[|0\rangle\otimes|-\rangle + (|2\rangle + |4\rangle)\otimes (|-\rangle + |+\rangle)\right].
\end{equation}
After the first step the system's state is
\begin{equation}\label{pre1}
|pre(1)\rangle = \frac{1}{\sqrt{5}}\left[(|1\rangle + |3\rangle)\otimes (|-\rangle + |+\rangle) + |5\rangle\otimes|+\rangle\right].
\end{equation}
and after the second step it is
\begin{equation}
|pre(2)\rangle = |pre(0)\rangle
\end{equation}

Once the two steps are implemented, we are going to measure if the system is in the state (post-select it)
\begin{equation}\label{post2}
|post(2)\rangle = \frac{1}{\sqrt{5}}\left[|0\rangle\otimes|-\rangle + (|2\rangle + |4\rangle)\otimes (|-\rangle - |+\rangle)\right].
\end{equation}
The probability of post-selection is
\begin{equation}
|\langle post(t)|pre(t)\rangle|^2=\frac{1}{25}
\end{equation}
Finally, notice that the post-selected state can be in principle evolved backwards in time
\begin{equation}
|post(1)\rangle = \frac{1}{\sqrt{5}}\left[(|1\rangle + |3\rangle)\otimes (|+\rangle - |-\rangle) + |5\rangle\otimes|+\rangle\right].
\end{equation}
\begin{equation}
|post(0)\rangle = |post(2)\rangle
\end{equation}


\subsection{Paradox}

Pre- and post-selection can lead to paradoxes \cite{PPS}, such as the three-box one \cite{3Box}.  In our case the paradox originates from the following counterfactual reasoning \cite{UsNJP}. Imagine that at time $t=0$ the system was pre-selected in the state $|pre(0)\rangle$ and at time $t=2$ it was post-selected in the state $|post(2)\rangle$. Now we ask: what if between pre-selection and post-selection somebody looked for the walker at a certain location? Interestingly, for some locations the answers are deterministic, yet counterintuitive (hence the name -- {\it logical} pre- and post-selection paradox \cite{LS1,LS2}). In particular, if at time $t=0$ (or $t=2$) one  asked if the walker is at position $x=0$, the answer would have to be YES. This is because if the answer were NO, the state $|pre(0)\rangle$ ($|pre(2)\rangle$) would collapse onto
\begin{equation}\label{preb0}
|pre_{\bar{0}}\rangle = \frac{1}{2}(|2\rangle + |4\rangle)\otimes (|-\rangle + |+\rangle).
\end{equation}
The collapsed state cannot be later measured as $|post(2)\rangle$, since $\langle pre_{\bar{0}}|post(2)\rangle=0$. Therefore, the answer would have to be YES due to the post-selection assumption. Similarly, if at time $t=1$ one asked if the walker is at position $x=5$, the answer would have to be YES too. If the answer were NO, the state $|pre(1)\rangle$ would collapse onto
\begin{equation}\label{preb5}
|pre_{\bar{5}}\rangle =  \frac{1}{2}(|1\rangle + |3\rangle)\otimes (|-\rangle + |+\rangle).
\end{equation}
which is orthogonal to $|post(1)\rangle$. We conclude that at $t=0$ and $t=2$ the walker was at $x=0$ and at $t=1$ it was at $x=5$. This is paradoxical, since the model allows the walker to jump only to neighbouring positions.


\subsection{Contextuality}

\begin{figure*}
  \centering

  \vspace{-0.5cm}
  \includegraphics[scale=0.5]{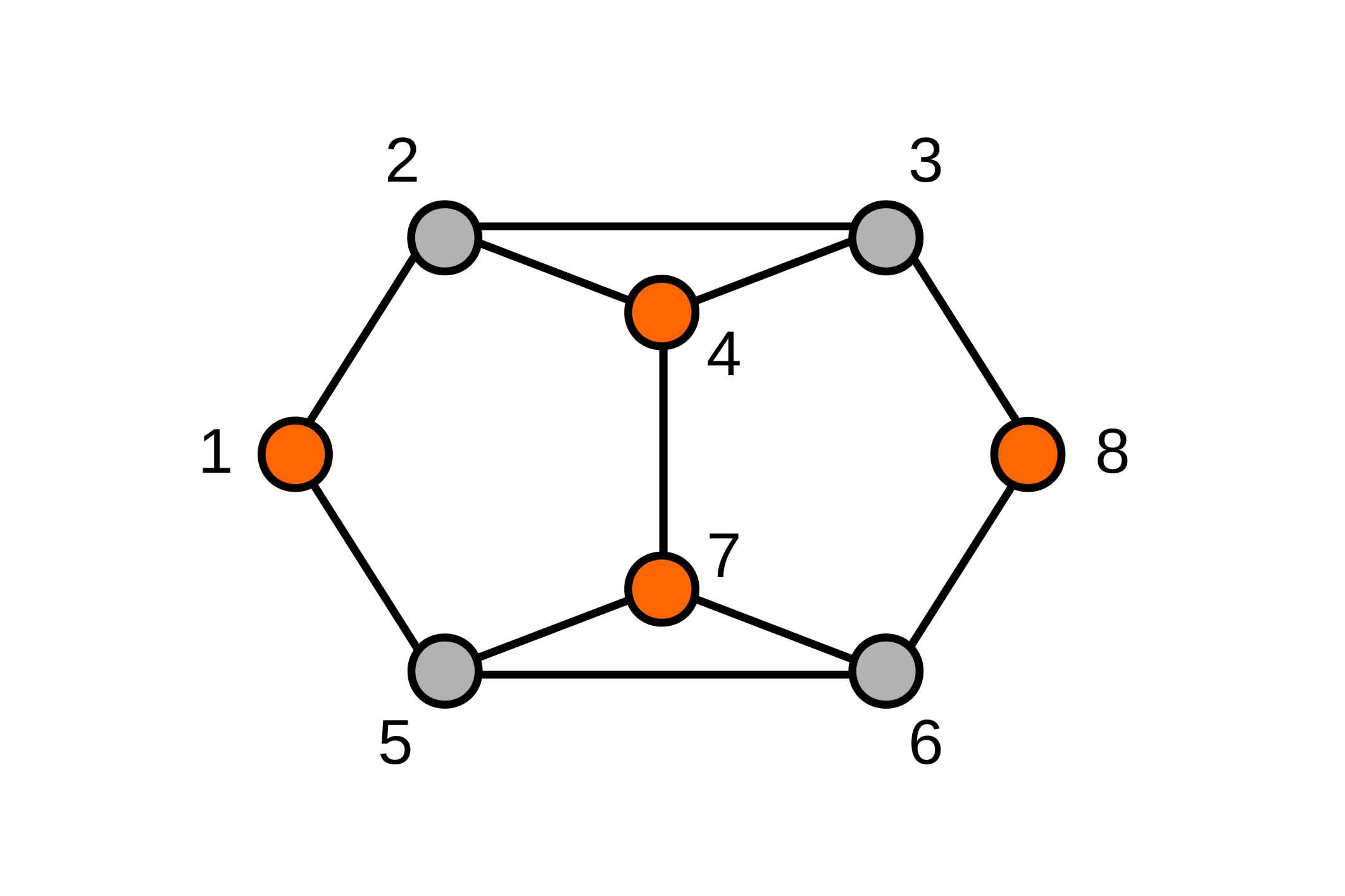}\includegraphics[scale=0.5]{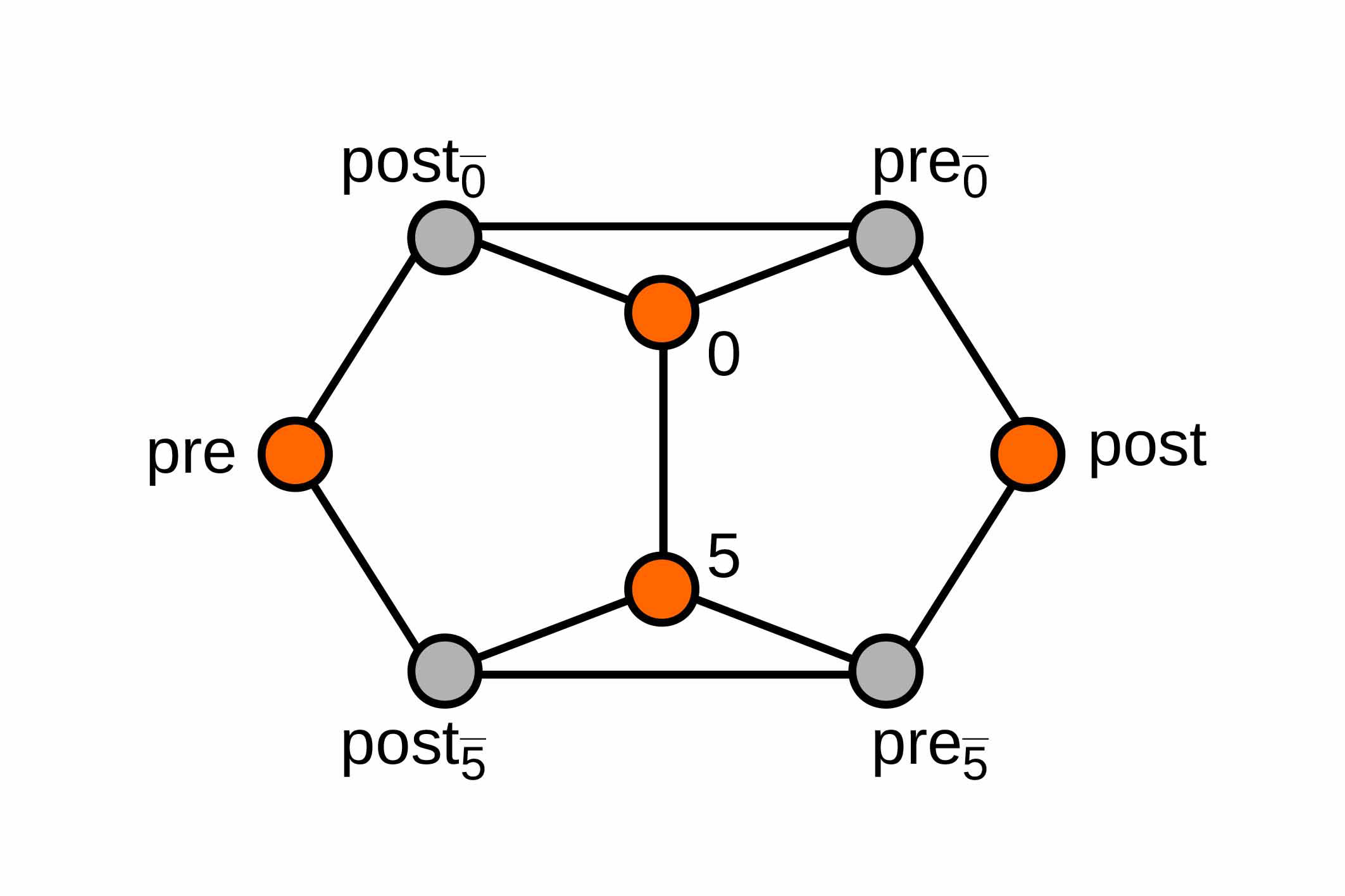}
  \vspace{-1cm}
  \caption{A graph representing the Clifton's proof of contextuality. The vertices represent measurable events and the edges represent the exclusivity relation between the events. The events are assigned logical values YES (orange) and NO (grey). In a NCHV theory the logical values are assigned to all events, i.e., all vertices need to be colored either orange or grey. The graph structure implies the paradox (explanation in the main text). Left -- the eight Clifton's events. Right -- the events in our quantum walk scenario.} \label{f1}
\end{figure*}

The above is a LPPS paradox since the probabilities of counterfactual events are either zero or one. Such paradoxes were shown to be equivalent to proofs of contextuality \cite{PPSc1,PPSc2}. Below we base on the work of Leifer and Spekkens \cite{LS1} and relate the quantum walk LPPS paradox to the Clifton's proof of contextuality \cite{Clifton}. The Clifton's proof can be formulated with a help of a graph -- see Fig. \ref{f1}. The vertices in the graph correspond to measurable events, whereas the edges denote the exclusivity relation between them. This means that if one event happens, any other event connected to it by an edge cannot happen. The NCHV theory assigns logical values YES and NO to all vertices (orange and grey color, respectively). This assignment represents an NCHV preparation of a system. The act of measurement in such a theory mearly reveals the pre-assigned values. The value YES means that the corresponding event will be observed (when measured), whereas NO means that it will not.

Let us first consider a general scenario of eight events (Fig. \ref{f1} left) $e=1,2,\ldots,8$.  We assume that the system is pre-selected in a state corresponding to the event 1 and post-selected in a state corresponding to the event 8. This means that within a NCHV theory the vertices 1 and 8 are assigned logical values YES. However, such assignment implies that the events 2 and 5 (exclusive to 1) and events 3 and 6 (exclusive to 8) must be assigned NO. Finally, consider two mutually exclusive events, 4 and 7, such that 4 is complementary to 2 OR 3 and 7 is complementary to 5 OR 6. This means that if 2 and 3 are assigned NO, then 4 must be assigned YES. Similarly, if 5 and 6 are assigned NO, then 7 must be assigned YES. We obtained a contradiction, since due to exclusivity 4 and 7 cannot be both assigned YES. Hence, the system cannot be described by a NCHV theory.

As shown by Clifton, it is possible to find a quantum system for which there exists a set of eight events whose exclusivity relations match the ones represented in the above graph. The event $i$ corresponds to a projector $\Pi_i$. If two events $i$ and $j$ are exclusive, the corresponding projectors are orthogonal $\Pi_i \Pi_j = 0$. This means that such quantum system does not admit a NCHV description.

Now we return to our quantum walk scenario. In particular, we are going to show that there are eight events in our system corresponding to the Clifton's events -- see Fig. \ref{f1} right. Clearly, the events 1 and 8 correspond to projectors
\begin{equation}
\Pi_{pre} = |pre(0)\rangle\langle pre(0)| = |pre(2)\rangle\langle pre(2)|
\end{equation}
and
\begin{equation}
\Pi_{post} = |post(0)\rangle\langle post(0)| = |post(2)\rangle\langle post(2)|
\end{equation}
The two exclusive events 4 and 7 correspond to quantum walk events that at time $t=0,2$ the particle is at $x=0$ and that at time $t=1$ the particle is at $x=5$, respectively. In Fig. \ref{f1} right these events are denoted as $0$ and $5$. They correspond to projectors
\begin{equation}
\Pi_0 = |0\rangle\langle 0|\otimes \openone_c
\end{equation}
and
\begin{equation}
\Pi_5 = U|5\rangle\langle 5|\otimes \openone_c U^{\dagger}
\end{equation}
where $\openone_c$ is the identity operator on the coin space and $U$ is the unitary operator generating a single step of the evolution. The operator $U$ is included in the projector $\Pi_5$ due to the fact that this measurement is done at time $t=1$, i.e., after one step of the evolution. Note that $U^2 = \openone$, hence the evolution operator does not need to be included in projectors corresponding to measurements done at $t=2$.

We are left with four Clifton's events: 2, 3, 5 and 6. The event 2, which we label $post_{\bar{0}}$, corresponds to a situation in which at time $t=0,2$ the particle's coin state is $|c_-\rangle \equiv \frac{1}{\sqrt{2}}(|-\rangle - |+\rangle)$ and it's position is not $x=0$. The associated projector is
\begin{equation}
\Pi_{post_{\bar{0}}}= (\openone_x -|0\rangle\langle 0|)\otimes |c_-\rangle\langle c_-|
\end{equation}
where $\openone_x$ is the identity operator in the position space. Note that this projector is orthogonal to both, $\Pi_{pre}$ and $\Pi_0$. In addition, since we consider a particular post-selection, effectively
\begin{equation}
\Pi_{post_{\bar{0}}} = |post_{\bar{0}}\rangle\langle post_{\bar{0}}|
\end{equation}
where
\begin{equation}
|post_{\bar{0}}\rangle = \frac{1}{2}(|2\rangle + |4\rangle)\otimes (|-\rangle - |+\rangle).
\end{equation}
Similarly, the event 5, which we label $post_{\bar{5}}$, corresponds to a situation in which at time $t=1$ the particle's coin state is $|c_-\rangle$ and it's position is not $x=5$. The associated projector is
\begin{equation}
\Pi_{post_{\bar{5}}}= U(\openone_x -|5\rangle\langle 5|)\otimes |c_-\rangle\langle c_-|U^{\dagger}
\end{equation}
It is orthogonal to both, $\Pi_{pre}$ and $\Pi_5$. Moreover, as before, due to post-selection effectively
\begin{equation}
\Pi_{post_{\bar{5}}} = U|post_{\bar{5}}\rangle\langle post_{\bar{5}}|U^{\dagger}
\end{equation}
where
\begin{equation}
U|post_{\bar{5}}\rangle = U \frac{1}{2}(|1\rangle + |3\rangle)\otimes (|+\rangle - |-\rangle).
\end{equation}
Next, the event 3, labelled $pre_{\bar{0}}$, corresponds to a situation in which at time $t=0,2$ the particle's coin state is $|c_+\rangle \equiv \frac{1}{\sqrt{2}}(|-\rangle + |+\rangle)$ and it's position is not $x=0$. The associated projector
\begin{equation}
\Pi_{pre_{\bar{0}}}= (\openone_x -|0\rangle\langle 0|)\otimes |c_+\rangle\langle c_+|
\end{equation}
is orthogonal to $\Pi_{post_{\bar{0}}}$, $\Pi_{0}$ and $\Pi_{post}$. This time, the pre-selection implies that effectively
\begin{equation}
\Pi_{pre_{\bar{0}}} = |pre_{\bar{0}}\rangle\langle pre_{\bar{0}}|
\end{equation}
where $|pre_{\bar{0}}\rangle$ is given in Eq. (\ref{preb0}). Finally, the event 6, labelled $pre_{\bar{5}}$, corresponds to a situation in which at time $t=1$ the particle's coin state is $|c_+\rangle$ and it's position is not $x=5$. The associated projector
\begin{equation}
\Pi_{pre_{\bar{5}}}= U(\openone_x -|5\rangle\langle 5|)\otimes |c_+\rangle\langle c_+|U^{\dagger}
\end{equation}
is orthogonal to $\Pi_{post_{\bar{5}}}$, $\Pi_{5}$ and $\Pi_{post}$. Once more, the pre-selection implies that effectively
\begin{equation}
\Pi_{pre_{\bar{5}}} = U|pre_{\bar{5}}\rangle\langle pre_{\bar{5}}|U^{\dagger}
\end{equation}
where $|pre_{\bar{5}}\rangle$ is given in Eq. (\ref{preb5}).

The above set of projectors, under the pre- and post-selection assumption, constitute a proof of contextuality in our quantum walk model. In the next section we are going to transform it into an experimentally testable inequality.


\subsection{Experimentally testable inequality}

The above proof of contextuality is a mathematical statement. It is not an experimentally realisable tests. To make it testable in a laboratory we transform it into an inequality that sets some non-contextual hidden variables (NCHV) bound on a function of measurable data. If the inequality is violated, the tested system is confirmed to be contextual.

We first reduce the Clifton's proof to the Wright/KCBS (Klyachko-Can-Binicioglu-Shumovsky) scenario \cite{Wright,KCBS}. The system is prepared in the state $|pre(0)\rangle$, but only five out of eight measurement events are considered: $pre_{\bar{0}}$, $0$, $5$, $pre_{\bar{5}}$, $post$. These five events form a subgraph -- a 5-cycle. The exclusivity relations in such a graph imply that at most two events can be assigned a logical value YES, hence in the NCHV theory the following inequality must hold
\begin{equation}\label{KCBS}
p(pre_{\bar{0}})+p(0)+p(5)+p(pre_{\bar{5}})+p(post) \leq 2.
\end{equation}
where $p(X)$ denotes the probability that an event $X$ happens.

Next, we take advantage of the fact that $\{pre_{\bar{0}}, 0,post_{\bar{0}}\}$ and $\{pre_{\bar{5}}, 5,post_{\bar{5}}\}$ form a complete set of mutually exclusive events, hence
\begin{eqnarray}
p(pre_{\bar{0}}) + p(0) = 1 - p(post_{\bar{0}}), \\
p(pre_{\bar{5}}) + p(5) = 1 - p(post_{\bar{5}}).
\end{eqnarray}
Plugging the above to (\ref{KCBS}) we get our final inequality
\begin{equation}\label{Ineq}
p(post) \leq p(post_{\bar{0}}) + p(post_{\bar{5}})
\end{equation}
This is an inequality that can be tested in laboratory. It's violation would indicate that our quantum walk model is contextual.

The pre-selection in the state $|pre(0)\rangle$ and post-selection in the state $|post(2)\rangle$ imply that one can evaluate the above probabilities in a relatively simple way. The probability $p(post)$ is simply the probability of post-selection. On the other hand, the probabilities $p(post_{\bar{0}})$ and $p(post_{\bar{5}})$ are the probabilities of post-selection in situations in which the positions 0 and 5 are blocked, respectively. The theoretical estimation of these probabilities yields violation
\begin{equation}\label{Ineq29}
\frac{1}{25} \leq 0 + 0.
\end{equation}
We confirmed this theoretical prediction in the single-photon quantum walk experimental scenario.


\subsection{Experimental Verification}

\begin{figure*}
	\centering
	\includegraphics[width=\linewidth]{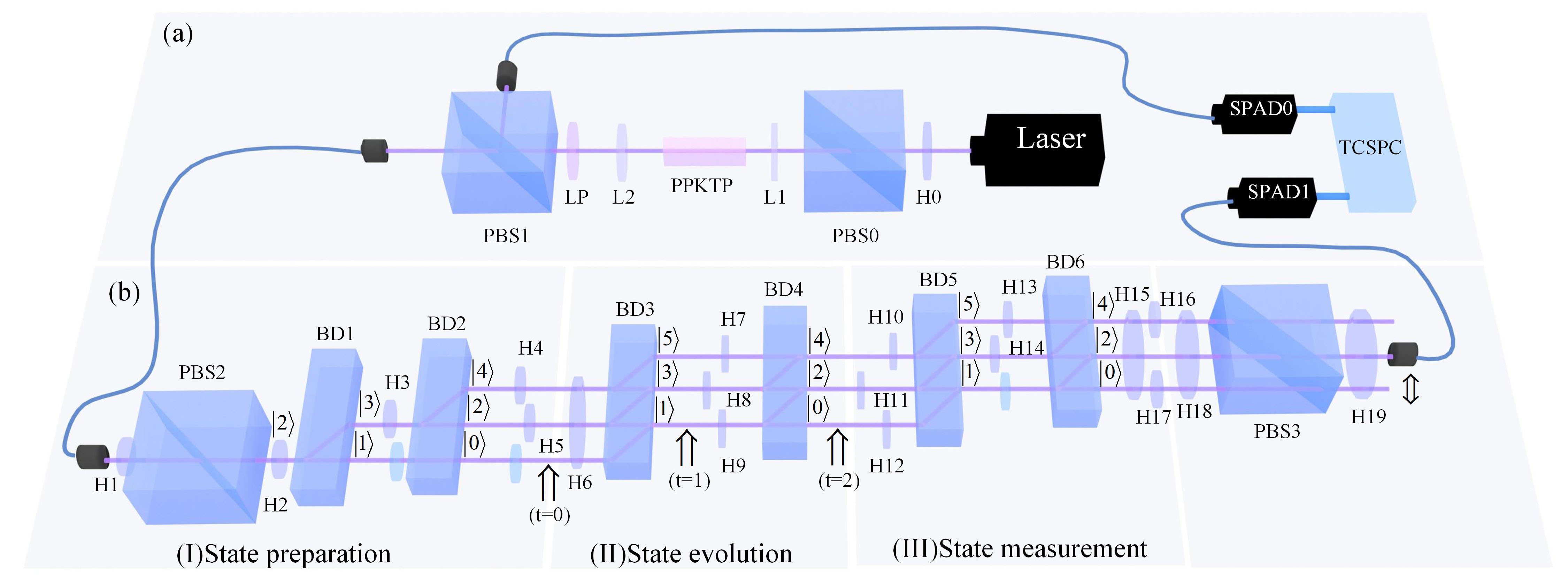}
\caption{Experimental setup. (a) Single photon source preparation. A 405nm continuous wave (CW) diode laser with 20mW power pumps a periodically poled potassium titanyI phosphate (PPKTP) crystal to produce photon pairs with central wavelength of 810 nm based on SPDC. The H0 and PBS0 are used to regulate optical power. The two lenses (L1 and L2) before and after PPKTP are used for focusing and collimating beams respectively. Then, after filtering out the pumped laser with a long pass filter (LP), the photon pairs are split on PBS1 and are coupled into single-mode fibers (SMF). (b) State preparation, state evolution and state measurement. The numbers on the figure represent the position of the walker. The unlabeled blue objects in the figure are all glass plates used to compensate for the phase. Ideally, after BD6, only a HWP is needed on the path $|2\rangle$. However, in the experiment we used HWP15 that spanned three paths. Therefore, HWP16 and HWP17 are needed to ensure that the polarization of $|0\rangle$ and $|4\rangle$ paths are in $|V\rangle$ mode. Finally, we detect photons at the transmission end of PBS3 by adjusting HWP18 and HWP19. The quantum walk photons are detected by SPAD1, whereas the heralding photon is detected by SPAD0. The effective coincidence window (including the jitter of the detector) is about 2ns.}
\label{fig:2}
\end{figure*}

\begin{figure}[htbp]
	\centering
    \includegraphics[width=\linewidth]{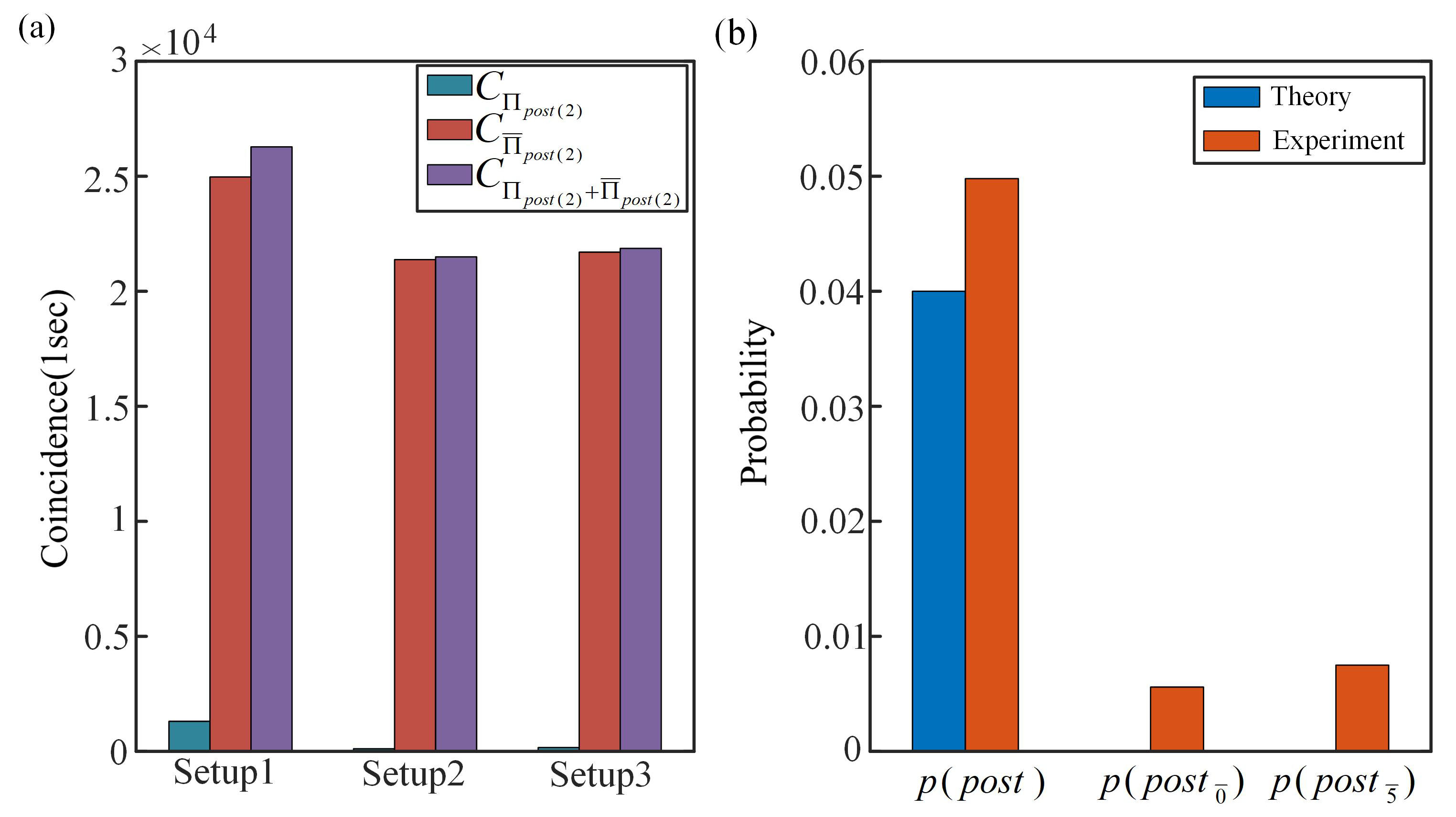}
	\caption{Experimental results. (a) Coincidence photon counts. The distribution of the number of photons measured post-selection in three setups. The x-coordinate represents the three experimental setups, and the y-coordinate represents the coincidence photon counts. Blue bars represent the coincidence photon counts between SPAD0 and the output of $\Pi_{post(2)}$. Red bars represent the coincidence photon counts between SPAD0 and the outputs of the complement of $\Pi_{post(2)}$, i.e., $\bar{\Pi}_{post(2)}$. Purple bars represent the total coincidence photon counts. (b) The probability distributions for the three setups. All of them are calculated from the raw data.}
\label{fig:5}
\end{figure}

We now demonstrate an experimental implementation of a one-dimensional DTQW and a violation of the inequality (\ref{Ineq}). The experimental setup is illustrated in Fig.\ref{fig:2}. The DTQW implementation is composed of three modules designed for state preparation, state evolution and state measurement (for more details see Methods). In order to demonstrate a violation of the inequality (\ref{Ineq}) we use three versions of the setup.

Setup 1: In the first setup there is no path blocked.
This allows us to evaluate the probability
\begin{equation}
p(post) =|\langle post(2)|pre(2)\rangle|^2.
\end{equation}

Setup 2: In the second setup the position $x=0$ is blocked at time $t=0$. Therefore, we can evaluate the probability
\begin{equation}
p(post_{\bar{0}}) =|\langle post(2)|pre_{\bar{0}}(2)\rangle|^2.
\end{equation}

Setup 3: In the third setup the position $x=5$ is blocked at time $t=1$. Hence, we can evaluate the probability
\begin{equation}
p(post_{\bar{5}}) =|\langle post(2)|pre_{\bar{5}}(2)\rangle|^2.
\end{equation}

The experimental results are shown in Fig.\ref{fig:5}. Fig.\ref{fig:5}(a) represents the coincidence photon counts. All the data shown in the figure is raw (without removing background noise). The post-selection measurement probabilities are shown in Fig.\ref{fig:5}(b) and the values obtained are
\begin{equation}
p(post) =\frac{C_{\Pi_{post(2)}}}{C_{\Pi_{post(2)}+\bar{\Pi}_{post(2)}}}=0.0498. ~~(Setup1)
\end{equation}
\begin{equation}
p(post_{\bar{0}}) = \frac{C_{\Pi_{post(2)}}}{C_{\Pi_{post(2)}+\bar{\Pi}_{post(2)}}}=0.0056.  ~~(Setup2)
\end{equation}
\begin{equation}
p(post_{\bar{5}}) = \frac{C_{\Pi_{post(2)}}}{C_{\Pi_{post(2)}+\bar{\Pi}_{post(2)}}}=0.0075.  ~~(Setup3)
\end{equation}
In the above $C_{\Pi_{post(2)}}$ denotes the number of coincidence counts for which the post-selection happened and $C_{\Pi_{post(2)}+\bar{\Pi}_{post(2)}}$ denotes the total number of coincidence counts.

It is obvious that these experimentally measured probabilities violate the inequality (\ref{Ineq})
\begin{equation}
p(post) = 0.0498 \nleq p(post_{\bar{0}}) + p(post_{\bar{5}})=0.0131
\end{equation}
It can be noted that the obtained experimental values are slightly higher than the theoretical ones. The difference is mainly due to a systematic error caused by the limited precision of wave plates and the imperfect visibility of Mach-Zehnder interferometers. Since the single photon detection efficiency is not high, we must adopt the fair sampling hypothesis, which is a standard assumption in experiments of this type.

\section{Summary}

We proposed a LPPS paradox based on a DTQW. Namely, we showed that under pre- and post-selection assumption a single-photon DTQW  undergoes a non-classical long distance oscillations. Next, we related this paradox to the Clifton's  proof of contextuality. Then, we transformed this proof into a new Bell-like inequality, whose violation confirms contextuality of the underlying system. Finally, we experimentally implemented a single-photon DTQW and demonstrated violation of the above inequality up to 8 standard deviations. Therefore, we proved for the first time that DTQWs cannot be described by a non-contextual hidden variable theory.


\section{Methods}

\subsection{Single photon generation}

\begin{figure}[htbp]
	\centering
	\includegraphics[width=\linewidth]{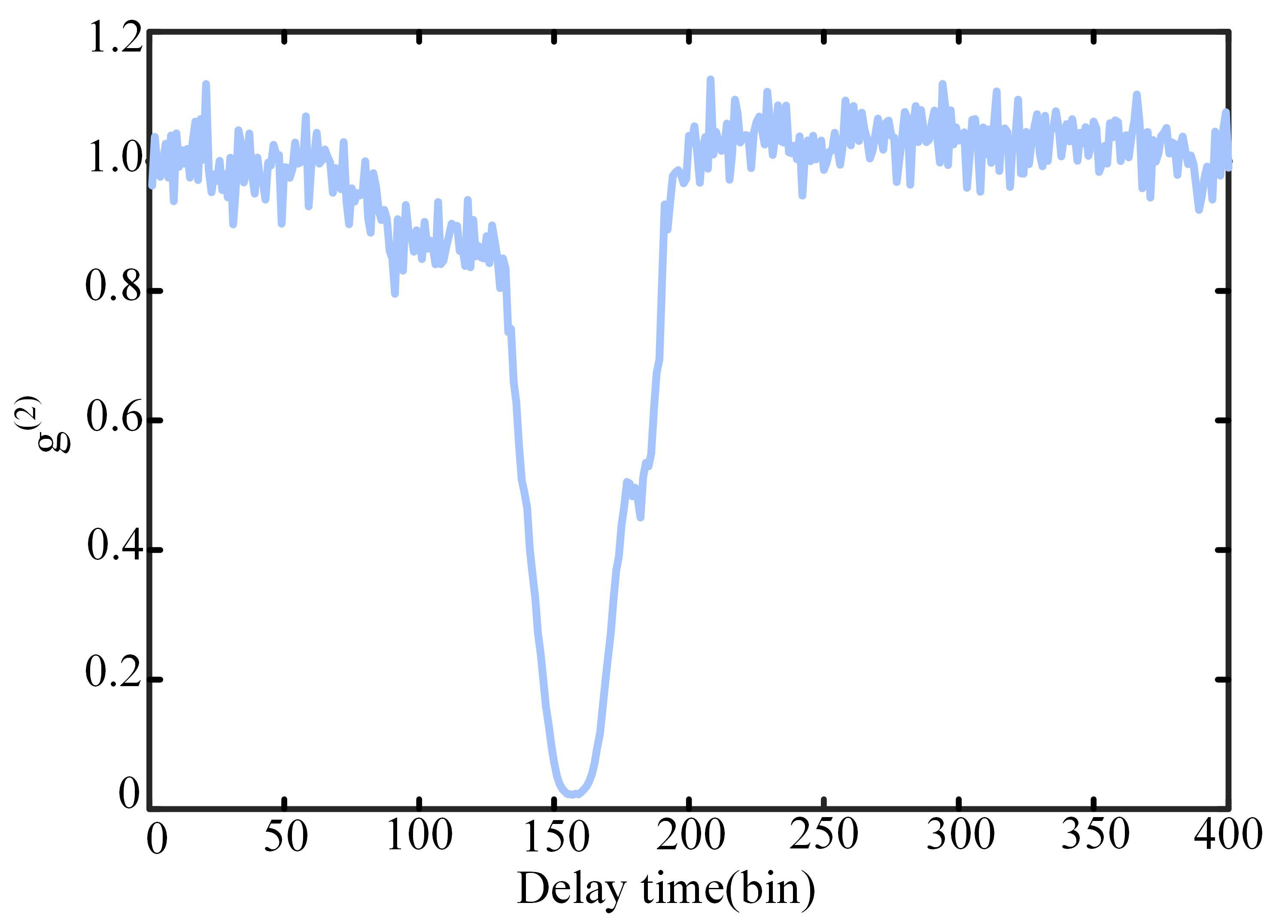}
\caption{Second-order correlation of a single-photon source. There is a clear dip in 160 time bins and every bin is 64ps.}
\label{fig:3}
\end{figure}

As shown in Fig.\ref{fig:2}(a), in the spontaneous parametric down-conversion (SPDC) the pairs of orthogonally polarized photons are produced in a polarization product state. We separate them at the polarizing beam splitter 1 (PBS1). A detection of a vertically polarized photon at SPAD0 heralds a horizontally polarized photon in our setup \cite{APL}. Similar to \cite{xiaoxiao}, we first conduct the Hanbury Brown-Twiss (HBT) experiment to confirm that the light source used in our experiment is a single-photon source. In Fig.\ref{fig:3} a clear dip represents the minimum value of $g^{(2)} = 0.0293$. This confirms that our source is indeed a single-photon source.

\subsection{Experimental DTQW setup}

The photonic DTQW setup consists of three parts, shown in Fig.\ref{fig:2}(b). In the state preparation part we use the half-wave plate 1 (HWP1, referred to as H1) and the PBS2 to regulate the light power and to ensure that the input light in the subsequent system is horizontally polarized. The two degrees of freedom, the single-photon polarization and the path, are used to encode the coin states $|\pm\rangle$ and the position states $|x\rangle$ of the walker, respectively. To prepare the initial state we use beam displacers (BDs) and HWPs. BDs transmit vertically polarized photons and displace horizontally polarized ones.
Initially the system's state is $|2\rangle\otimes|H\rangle$. After HWP2-HWP5, BD1 and BD2 it becomes
\begin{equation}\label{p1}
|pre(0)\rangle = \frac{1}{\sqrt{5}}\left[|0\rangle\otimes|V\rangle + (|2\rangle + |4\rangle)\otimes (|V\rangle + |H\rangle)\right].
\end{equation}
The above state corresponds to (\ref{pre0}). This ends the preparation (pre-selection) stage.

In the state evolution part the unitary operator generating a single step is realized via a combination of HWPs and BDs. The elements H6 and BD3 evolve the system from $t=0$ to $t=1$. The resulting state is
\begin{equation}
|pre(1)\rangle = \frac{1}{\sqrt{5}}\left[(|1\rangle + |3\rangle)\otimes (|V\rangle + |H\rangle) + |5\rangle\otimes|H\rangle\right],
\end{equation}
which corresponds to (\ref{pre1}). Similarly, HWP7-HWP9 and BD4 are used to evolve the system from $t=1$ to $t=2$. The system's state returns to $|pre(0)\rangle$ (since $|pre(2)\rangle=|pre(0)\rangle$).

\begin{figure}[htbp]
	\centering
	\includegraphics[width=\linewidth]{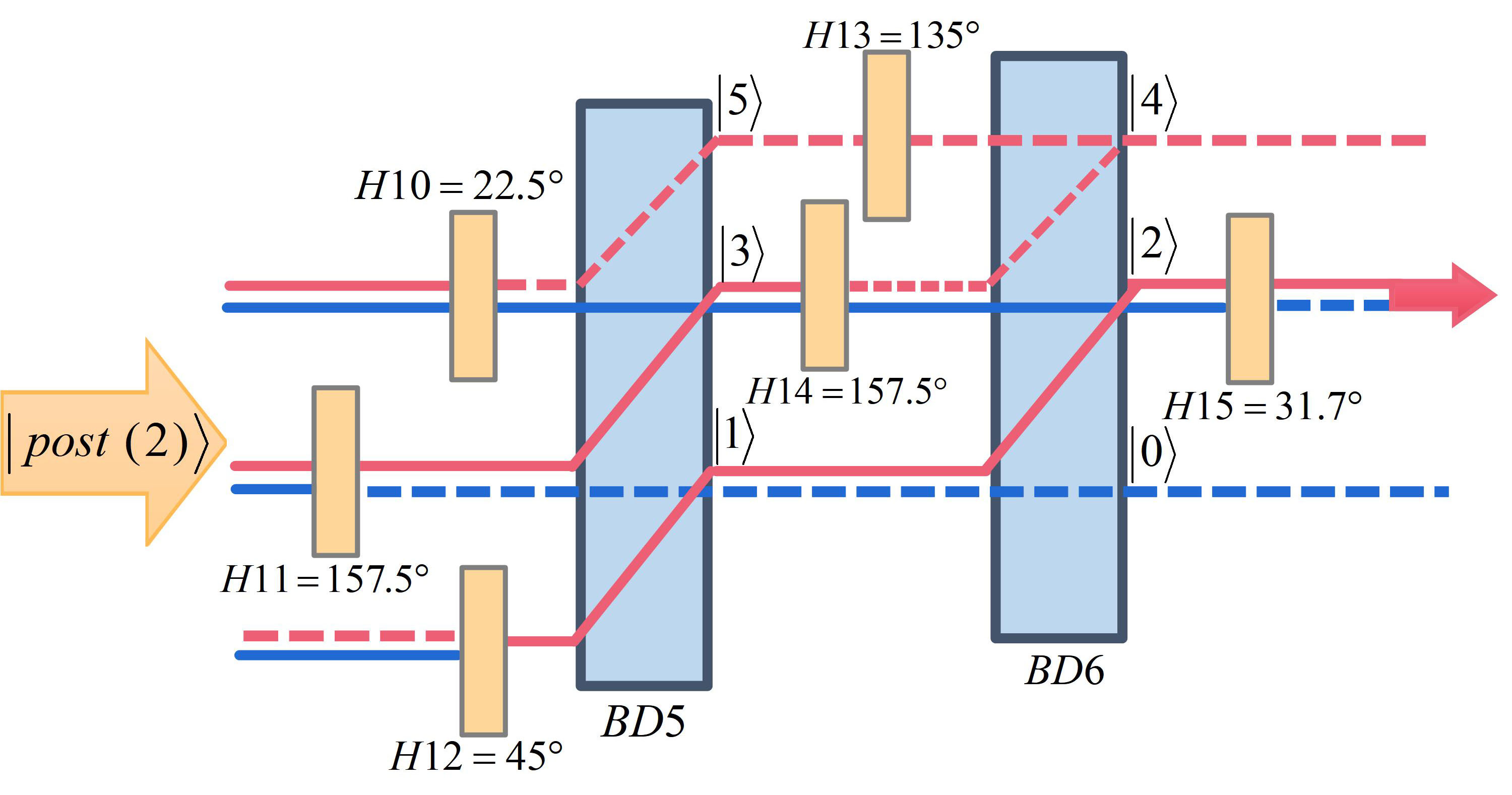}
	\caption{Experimental setup for measuring single observable. $|post(2)\rangle$ ($|post(\bar{2})\rangle$, which represents the orthogonal of $|post(2)\rangle$) is the eigenstate of $\Pi_{post(2)}$ corresponding to the eigenvalue -1(+1). The number accompanying each HWP is the angle of its optical axis relative to the horizontal polarization direction. The red line represents the $|H\rangle$ polarization in the path, and the blue line represents the $|V\rangle$ polarization. The red(blue) dotted line shows that there is no photon in the $|H\rangle$ polarization($|V\rangle$ polarization) of this path.}
\label{fig:4}
\end{figure}

In the last part we measure (post-select) the system in the state
\begin{equation}
|post(2)\rangle = \frac{1}{\sqrt{5}}\left[|0\rangle\otimes|V\rangle + (|2\rangle + |4\rangle)\otimes (|V\rangle - |H\rangle)\right]
\end{equation}
that corresponds to (\ref{post2}). Therefore, we need to design a proper measurement device. According to \cite{Huang}, it is a typical single-observable measuring devices (see Fig.\ref{fig:4}). The settings of HWP10-HWP15 are chosen to transform the state $|post(2)\rangle$  onto $|2\rangle\otimes |H\rangle$. The angles of all the HWPs used above are shown in the TABLE \ref{TableI}. Finally, output photons are detected using single-photon detector that consists of single-photon avalanche photo-diode (SPAD) and time-correlated single-photon counting (TCSPC). We register coincidences between SPAD1 (D1) and the trigger SPAD0 (D0). For each measurement, we record clicks for 1 s, registering about 11000 heralded single photons.

\begin{table}
\caption{Setting angles of the wave plates for realizing the total experimental setup}
\centering
\begin{tabular}{lcccccccr}
\hline
\hline
  HWP & HWP2 & HWP3 & HWP4 & HWP5\\
\hline
  $\theta$ & $13.3^{\circ}$ & $157.5^{\circ}$ & $22.5^{\circ}$ & $157.5^{\circ}$ \\
\hline
  HWP & HWP6 & HWP7 & HWP8 & HWP9\\
\hline
  $\theta$ & $45^{\circ}$ & $45^{\circ}$ & $45^{\circ}$ & $45^{\circ}$\\
\hline
  HWP & HWP10 & HWP11 & HWP12 & HWP13\\
\hline
  $\theta$ & $22.5^{\circ}$ & $157.5^{\circ}$ & $45^{\circ}$ & $135^{\circ}$ \\
\hline
  HWP & HWP14 & HWP15 & HWP16 & HWP17\\
\hline
  $\theta$ & $157.5^{\circ}$ & $31.7^{\circ}$ & $-58.3^{\circ}$ & $-58.3^{\circ}$\\
\hline
\hline
\end{tabular}
\label{TableI}
\end{table}


\section*{Acknowledgements}

This work is supported by National Key R\&D Program of China(2018YFB0504300). At the early stage of research TK and PK were supported by the Ministry of Science and Higher Education in Poland (science funding scheme 2016-2017 project no. 0415/IP3/2016/74). At the final stage of research PK was supported by the Polish National Science Centre (NCN) under the Maestro Grant no. DEC-2019/34/A/ST2/00081.



\begin{thebibliography}{99}

\bibitem{Aharonov}
Y. Aharonov, L. Davidovich, N. Zagury, Phys. Rev. A {\bf 48}, 1687 (1993).

\bibitem{Meyer}
D. A. Meyer, J. Stat. Phys. {\bf 85}, 551 (1996).

\bibitem{Review1}
J. Kempe, Cont. Phys. {\bf 44}, p.307-327 (2003).

\bibitem{Review2}
V. Kendon, Math. Struct. in Comp. Sci {\bf 17}, 1169-1220 (2006).

\bibitem{Review3}
D. Reitzner, D. Nagaj, V. Buzek, Acta Phys. Slov. {\bf 61}, 603-725 (2011).

\bibitem{Review4}
S. E. Venegas-Andraca, Quant. Inf. Proc. {\bf 11}, 1015-1106 (2012).

\bibitem{CQW1}
P. L. Knight, E. Roldan and J. E. Sipe, Phys. Rev. A {\bf 68}, 020301 (2003).

\bibitem{CQW2}
H. Jeong, M. Paternostro and M. S. Kim, Phys. Rev. A {\bf 69}, 012310 (2004).

\bibitem{CQW3}
S. K. Goyal, F. S. Roux, A. Forbes and T. Konrad, Phys. Rev. Lett. {\bf 11}, 0263602 (2013).

\bibitem{CQW4}
B. Sephton {\it et. al.}, PLoS One {\bf 14}, e0214891 (2019).

\bibitem{Fine}
A. Fine, Phys. Rev. Lett. {\bf 48}, 291 (1982).

\bibitem{Bell}
J. S. Bell, Physics {\bf 1}, 195 (1964).

\bibitem{KS}
S. Kochen and E. P. Specker, J. Math. Mech. {\bf 17}, 59 (1967).

\bibitem{LG}
A. J. Leggett and A. Garg, Phys. Rev. Lett. {\bf 54}, 857 (1985).

\bibitem{QWLG}
C. Robens, W. Alt, D. Meschede, C. Emary and A. Alberti, Phys. Rev. X {\bf 5}, 011003 (2015).

\bibitem{UsNJP}
T. Kopyciuk, M. Lewandowski, P. Kurzy{\'n}ski, New J. Phys. {\bf 21}, 103054 (2019).

\bibitem{PPSc1}
M. F. Pusey, Phys. Rev. Lett. {\bf 11}, 3200401 (2014).

\bibitem{PPSc2}
M. F. Pusey and M. S. Leifer, Proc. QPL 2015, EPTCS vol 195, p295 (2015).

\bibitem{PPS}
Y. Aharonov, P. G. Bergmann and J. L. Lebowitz, Phys. Rev. {\bf 134} B1410 (1964).

\bibitem{3Box}
Y. Aharonov and L. Vaidman, J. Phys. A: Math. Gen. {\bf 24} 2315 (1991).

\bibitem{LS1}
M. S. Leifer and R. W. Spekkens, Phys. Rev. Lett. {\bf 95}, 200405 (2005).

\bibitem{LS2}
M. S. Leifer and R. W. Spekkens, Proceedings of QS 2004, Int. J. Theor. Phys. {\bf 44}, 1977 (2005).

\bibitem{Wright}
R. Wright, In Mathematical Foundations of Quantum Mechanics,
edited by A. R. Marlow (Academic Press, San Diego), p. 255, (1978).

\bibitem{KCBS}
A. A. Klyachko, M. A. Can, S. Binicioglu and A. S. Shumovsky, Phys. Rev. Lett. {\bf 101} 020403 (2008)

\bibitem{Clifton}
R. Clifton, Am. J. Phys. {\bf 61}, 443 (1993).

\bibitem{APL}
J. Z. Yang, M. F. Li, X. X. Chen, W. K. Yu and A. N. Zhang,  Appl. Phys. Lett. {\bf 117} 214001 (2020)

\bibitem{xiaoxiao}
X. X. Chen, J. Z. Yang, X. D. C and A. N. Zhang, Phys. Rev. A {\bf 100} 042302 (2019).

\bibitem{Huang}
Y. F. Huang, M. Li, D. Y. Cao, C. Zhang, Y. S. Zhang, B. H. Liu, C. F. Li and G. C. Guo, Phys. Rev. A {\bf 87} 052133 (2013).


\end{thebibliography}
\end{document}